\documentclass[prx,twocolumn,english,superscriptaddress,floatfix,longbibliography]{revtex4-2}
\usepackage{graphicx,amsmath,amssymb,natbib,color,soul,easyReview,comment,lineno,appendix}

\usepackage{dcolumn}
\usepackage{bm}

\usepackage[utf8]{inputenc}
\usepackage[T1]{fontenc}
\usepackage{mathptmx}
\usepackage{etoolbox}
\usepackage{hyperref}

\newcommand{\rev}[1]{{\textcolor{black}{#1}}}

\begin{document}

\preprint{AIP/123-QED}

\title{Starting from the amorphous ground state: linking landscape thermodynamics to slow dynamics and crossover}
 
\author{Anshul D. S. Parmar}
\email{aparmar@uni-muenster.de}
\affiliation{\textit{Institute of Physical Chemistry, University of M\"{u}nster, Corrensstra{\ss}e 28/30, 48149 M\"{u}nster, Germany}}
\affiliation{\textit{Helmholtz-Institute M\"{u}nster (IEK-12) Forschungszentrum J\"{u}lich GmbH, Corrensstra{\ss}e 46, 48149 M\"{u}nster, Germany}}

\author{Simon G. Kellers}
\affiliation{\textit{Institute of Physical Chemistry, University of M\"{u}nster, Corrensstra{\ss}e 28/30, 48149 M\"{u}nster, Germany}}

\author{Andreas Heuer}
\email{andheuer@uni-muenster.de}
\thanks{\\This article may be downloaded for personal use only. Any other use requires prior permission of the author and AIP Publishing. This article appeared in The Journal of Chemical Physics and may be found at \url{https://doi.org/10.1063/5.0325432}.}
\affiliation{\textit{Institute of Physical Chemistry, University of M\"{u}nster, Corrensstra{\ss}e 28/30, 48149 M\"{u}nster, Germany}}

\begin{abstract}
{\rev{ 
A microscopic understanding of low-temperature thermodynamics and its relation to dynamical features such as a fragile-to-strong crossover (FSC) remains a central challenge in glass physics. Using swap Monte Carlo combined with a full potential-energy-landscape (PEL) analysis of a non-network-forming model, we obtain equilibrium data deep into the glassy regime and identify a finite system size that simultaneously reproduces bulk behaviour for $T \gtrsim T_g/2$ and allows complete sampling of the PEL down to its lowest-energy amorphous states. This enables the direct computation of the configurational entropy over the full temperature range of the finite system without relying on liquid-state thermodynamic integration. We find a pronounced depletion of low-energy states relative to the Gaussian regime of the PEL, which governs the low-temperature curvature of the configurational entropy. Numerically, the apparent activation energy of the diffusivity closely follows the temperature dependence of the mean inherent structure energy and exhibits a gradual crossover towards Arrhenius-like behaviour. This correlation is consistent with a trap-model description of the PEL, in which the FSC emerges naturally as a consequence of the depletion of low-energy states and thus of the lower bound of the PEL. We further argue, as illustrated analytically for a simple binomial model of the PEL, that the observability of a FSC depends on whether the depletion regime is reached within the accessible temperature window.
}}
\end{abstract}

\maketitle

\section{INTRODUCTION } 

Glass, a disordered material, has been used for millennia, yet the fundamental nature of the glass state and the processes leading to its formation remain unresolved~\cite{anderson1995through}. Owing to their broad relevance from - polymers and alloys to colloids - glass forming liquids appear in a wide variety of settings~\cite{debenedetti2001supercooled}. Experimentally, supercooling is possible over an equilibrium window spanning relaxation times from $\sim 10^{-10}$ to $10^{2}$\,s, defining the onset temperature $T_{onset}$ and the calorimetric glass transition temperature $T_g$~\cite{schmidtke2012boiling}.  

Dynamically, glass formers are broadly classified as \emph{strong} or \emph{fragile}, depending on whether their relaxation times obey an Arrhenius or super-Arrhenius temperature dependence~\cite{angell1991relaxation}. Some network-forming liquids, such as silica and water, display a fragile-to-strong crossover (FSC) upon cooling~\cite{saika2001fragile,saksaengwijit2004origin}. For such systems, the FSC correlates with features in the entropy curve~\cite{saika2001fragile}, suggesting a deep connection between thermodynamics and dynamics that, however, remains only partially understood.

Thermodynamically, relations between dynamics and entropy have been formalized in classical approaches such as Adam–Gibbs~\cite{adam1965temperature} and {\rev {Random First Order Transition}} (RFOT) ~\cite{lubchenko2006theory,kirkpatrick1989scaling}, and have been examined through several model-based frameworks~\cite{ruocco2004landscapes,parisi2020theory}. Nevertheless, intrinsic features of specific systems have made it difficult to establish a general, quantitative theory of the glass transition that simultaneously accounts for thermodynamic anomalies (entropy, landscape structure) and dynamical crossovers (FSC, deviations from simple super-Arrhenius behaviour).

{\rev{A natural route toward such a theory is provided by the potential energy landscape (PEL), which represents the system’s potential energy as a function of all particle coordinates. The landscape is partitioned into basins separated by energy barriers and saddle points; local minima (inherent structures, IS) correspond to mechanically stable configurations, where each basin configuration relaxes to that local minimum.}} ~\cite{goldstein1969viscous,angell1991transport,sastry2001relationship,sciortino2005potential,la2006relation,heuer2008exploring,baity2021revisiting}. 
{\rev{The distribution and topology of inherent structures encode key thermodynamic and dynamical properties. If the \emph{complete} density of states were known, including its lowest-energy states, one could in principle compute central thermodynamic quantities—such as entropy or specific heat—at all temperatures, where direct simulations are impossible due to astronomical relaxation times~\cite{berthier2011theoretical}. This framework provides insight into the origin of fragility, the concept of an ideal glass~\cite{stillinger1995topographic}, which in the Stillinger picture is associated with the lowest-energy amorphous states (deepest minima of the landscape). In contrast, within RFOT theory, the ideal glass is defined as a thermodynamic transition at the Kauzmann temperature $T_K$, where the configurational entropy extrapolates to zero. These viewpoints highlight that the notion of an ideal glass depends on the theoretical framework adopted.
In this work, we adopt the PEL perspective and use the term in this restricted sense.}}
Moreover, interpretations of the PEL in terms of a trap model have suggested that activated dynamics can be expressed directly in terms of landscape properties~\cite{heuer2008exploring}, but a fully quantitative, general link between thermodynamic landscape observables and dynamical activation energies has remained elusive.

Meaningful PEL analyses require small system sizes, because large systems split into many weakly correlated subsystems~\cite{doliwa2003finite,heuer2008exploring}. One therefore seeks the smallest system size $N_c$ that still reproduces bulk thermodynamic, structural, and dynamical behaviour. Reported values include $N_c=65$ for the 3D Kob–Andersen mixture~\cite{kob1994scaling,doliwa2003finite}, $N_c=99$ for 3D BKS silica~\cite{saksaengwijit2004origin}, and $N_c=80$ for 2D silica~\cite{roy2022influence}. For network formers, complete IS-densities have been resolved, exhibiting a Gaussian regime with a sharp depletion of low-energy states~\cite{saksaengwijit2004origin,roy2022influence}. This depletion has been connected to low-energy defect-free structures~\cite{saksaengwijit2004origin} and subtle structural changes~\cite{yu2022understanding}, and the associated cutoff explains the FSC within those models. However, beyond network formers, the complete IS-density for $N \approx N_c$ has remained unknown due to the exponentially growing computational cost at low energies, and even for network formers there is no quantitative characterization of the ideal glass or of a general, model-independent thermodynamics–dynamics relation.

Here we study a 2D non-network glass former using swap Monte Carlo~\cite{ninarello2017models}. Overcoming previous sampling limitations, we acquire equilibrium configurations for sizes $N\in[33,1056]$ down to $T \gtrsim T_g/2$ - far below the temperatures accessible in conventional simulations—without crystallization (supplementary material (SM), Sec.~V). This allows us to reconstruct the IS-density over the entire thermodynamic range relevant for deeply supercooled liquids. Crucially, after identifying $N_c$, we show that for all $N\le N_c$ the complete IS-density, including the lowest-energy states, can be sampled exhaustively. Consequently, for these finite system sizes, thermodynamic and structural quantities can be computed \emph{for all} temperatures $T>0$ (neglecting quantum effects), without relying on liquid-state entropy or thermodynamic integration. For $N>N_c$, insights from $N=N_c$ remain valid for the experimentally relevant regime $T\gtrsim T_g/2$.
{Furthermore, under the assumption of a mapping of the PEL on a trap model, we provide an exact identity, which links the apparent activation energy of the dynamics directly to the thermodynamic average IS energy.} 
 {This relation shows that deviations from Gaussian PEL behaviour, arising from the depletion of low-energy states, are {\rev{associated with}} the temperature dependence of the activation energy, and thus {\rev{provide a possible physical interpretation}} of the FSC. Using our fully resolved PEL for $N_c$, we employ this identity to quantify the size of the elementary relaxing subsystem and to demonstrate that, in our model, the FSC {\rev{is correlated with}} the same depletion regime that influences the configurational entropy at low temperatures. Finally, we {\rev{suggest}} that depletion at the bottom of the landscape, and its associated thermodynamics–dynamics coupling, {\rev{may represent}} a general and experimentally relevant mechanism of glass formation.}

\section{Model and Methods}

\subsection*{In-silico glass-forming model}

We study a discrete-polydisperse mixture of $N$ purely repulsive particles composed of $M_s=10$ species of equal mass $m$. Particle pairs $(i,j)$ interact via a truncated and smoothed power-law potential,
\begin{equation}
e(r_{ij})=\epsilon\left(\frac{\sigma_{ij}}{r_{ij}}\right)^{12}
+\sum_{k\in\{0,2\}}c_{2k}\left(\frac{r_{ij}}{\sigma_{ij}}\right)^{2k},
\end{equation}
with cutoff $r_{ij}/\sigma_{ij} \le 1.25$.  
The quartic correction ensures continuity of potential, force, and second derivative at the cutoff. We use $\epsilon$ as the unit of energy and temperature, the mean particle diameter $\langle\sigma\rangle$ as the unit of length, and reduced MD time units $\sqrt{m\langle\sigma\rangle^2/\epsilon}$.

Particle diameters are drawn from a distribution $P(\sigma)\propto \sigma^{-3}$ with $\sigma\in[0.73,1.62]$~\cite{ninarello2017models}.  
To enhance glass-forming ability and suppress crystallization, we employ non-additive mixing, $\sigma_{ij} = \tfrac{1}{2}(\sigma_i+\sigma_j)\bigl(1-0.2|\sigma_i-\sigma_j|\bigr)$. All simulations are performed in a square box of area $V$ at number density $\rho=N/V=1$ under periodic boundary conditions. 
This setup enables rigorous comparison across system sizes while maintaining identical disorder statistics~\cite{parmar2020ultrastable}.

\subsection*{Sample preparation and dynamics}

Equilibrium configurations in the deeply supercooled regime are generated using swap Monte Carlo~\cite{ninarello2017models}.  
One Monte Carlo step consists of 80\% translational moves and 20\% swap moves, and the Metropolis criterion ensures detailed balance.  
Extensive tests confirm the absence of aging and ensure true equilibrium sampling (SM, Sec.~I).

Local dynamics is obtained from molecular dynamics simulations using the No\'{s}e–Hoover thermostat in the range $T\in[0.5000,0.0690]$.  
For each system size $N\in[33,1056]$, we average over $n_s\in[100,1056]$ independent samples.

To suppress long-wavelength Mermin–Wagner fluctuations, which do not reflect cage-breaking processes~\cite{illing2017mermin}, we monitor dynamics via cage-relative coordinates,
\begin{equation}
\overline{\mathbf r}_i(t) = \mathbf r_i(t)
-\frac{1}{\mathcal N_i} \sum_{j\in\mathrm{n.n.}(i)} \mathbf r_j(t),
\end{equation}
where nearest neighbours are defined by $r_{ij}/\sigma_{ij}\le 1.33$.  
The diffusion constant is obtained from the long-time limit of the cage-relative mean-squared displacement,
\begin{equation}
\Delta(t)=\frac{1}{N}\Big\langle \sum_{i=1}^N  
\left[\overline{\mathbf r}_i(t)-\overline{\mathbf r}_i(0)\right]^2\Big\rangle.
\end{equation}

At high temperature the system displays simple-liquid behaviour obeying an Arrhenius law
\begin{equation}
D^{-1}(T)=D_\infty^{-1}\exp(-\beta A),
\end{equation}
with onset temperature $T_{onset}=0.25$.  
The mode-coupling temperature $T_{MCT}=0.105$ is obtained from the standard power-law fit  
$D^{-1}(T)\propto|T-T_{MCT}|^{-\gamma}$ with $\gamma=2.5$.  
The first three decades of glassy slowing down are described by parabolic~\cite{} or {\rev{Vogel–Fulcher–Tammann (VFT) forms}}, while the four–five lowest-temperature decades seem to exhibit an Arrhenius-type behaviour (SM, Sec.~III). Later it will be shown, that the low-temperature Arrhenius-behavior needs a more careful analysis.
The experimental glass transition $T_g$ corresponds to twelve decades of relaxation; our simulations for dynamics reach down to $\sim 1.35\,T_g$.

\subsection*{Potential energy landscape}

The potential energy landscape is characterized via inherent structures obtained by conjugate-gradient minimization to machine precision using \texttt{LAMMPS}~\cite{plimpton1995fast}.  
For each $N\in[33,1056]$, we collect $5\times10^5$–$10^7$ minimized configurations to construct the full energy distribution.

The Boltzmann distribution of IS energies satisfies
\begin{equation}
P(E_{IS},T) \propto G(E_{IS})\, e^{-\beta E_{IS}},
\end{equation}
so the volume-weighted IS-density $G(E_{IS})$ can be reconstructed via reweighting,
\begin{equation}
 P_W(E_{IS},T) \propto P(E_{IS},T)\, e^{\beta E_{IS}}.
 \label{eqn:PwIS_rew}
\end{equation}
Reweighted distributions are stitched together over the interval  
$\langle E_{IS}\rangle \pm 1.2\sigma$ (with $\sigma$ the IS-energy standard deviation).  
We restrict attention to the supercooled regime $T<0.25~(T_{onset})$, where the PEL displays a pronounced Gaussian regime at intermediate energies.

To validate equilibrium sampling, we investigate
(i) absence of aging and  
(ii) temperature-independence of the ratio 
$P_W(E_{IS},T_i) / P_W(E_{IS},T_j)$ across the studied temperature range (SM, Sec.~I).

\subsection*{Superposition hypothesis}

For sufficiently large systems we test whether an IS distribution for size $N$ can be described as the convolution of two independent subsystems of size $N/2$, i.e.

\begin{equation}
P(E_{IS};N)
= \int dE\, P(E;N_i)\,P(E_{IS}-E;N-N_i).
\label{eq:superposition}
\end{equation}

This superposition property becomes crucial for identifying the smallest system size $N_c$ that reliably reflects bulk thermodynamics, dynamics, and the structure of the PEL.

\subsection*{Structural disorder}

To quantify local structural fluctuations we use the disorder measure $\Theta$ introduced in Ref.~\cite{tong2018revealing}.  
For particle $i$ with Voronoi neighbours $\langle j,k\rangle$, we define
\begin{equation}
\Theta_i=\frac{1}{\mathcal N_i}
\sum_{\langle j,k\rangle}|\theta_{j,k}-\theta^{0}_{j,k}|,
\end{equation}
where $\theta_{j,k}$ is the angle at particle $i$ between neighbours $j,k$ in the IS configuration, and $\theta^{0}_{j,k}$ is the corresponding angle when all particles are in mutual contact, and $\mathcal N_i$ is the total number of neighbouring-pairs of particle $i$. 
The average disorder $\Theta$ provides a structural descriptor that correlates closely with IS energy in our model.


\begin{figure}[]
    \includegraphics[scale=1]{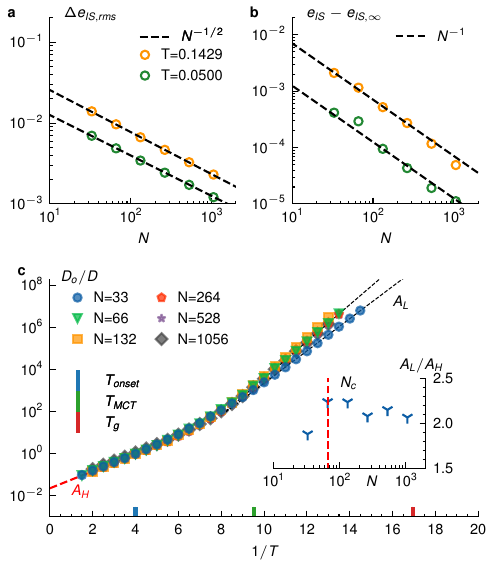}
    \caption{
    {(a)} The energy fluctuation per particle for two different temperatures. It decreases proportionally to $N^{-1/2}$.
    {\bf b} The finite-size effect of the average IS energy relative to the fluctuations, for the same two temperatures. Included is an empirical  $1/N$ asymptotic behavior (dashed line). 
    {(c)} The inverse diffusion constant $1/D$ (scaled with $D_o$, diffusion at the onset temperature) covers 7 decades of glassy dynamics. It reveals the crossover from fragile to strong behavior at low T($\lesssim 0.1$). The inset presents the low and high-T activation energy ratio for all system sizes.}\label{fig1}
\end{figure}

\section{Results and Discussion}

\subsection{Quantifying finite-size effects} 

We begin by characterizing how the IS   energy distribution depends on system size.  
For large systems, a configuration of size $2N$ can be regarded as a superposition of two weakly coupled subsystems of size $N$.  
This \emph{superposition hypothesis} implies that energy fluctuations per particle should scale as $N^{-1/2}$, since independent subsystems contribute additively.  
Figure~\ref{fig1}a confirms this prediction over the full range of $N$ studied.

The average IS energy per particle also displays a smooth size dependence.  
Empirically, the finite-size correction at fixed temperature is well described by  
$e_{IS}(N) - e_{IS}(N\!\to\!\infty) \approx 2c/N$,  
implying  
$N e_{IS}(N) - (2N) e_{IS}(2N) \approx c$.  
This reflects the fact that larger amorphous systems can more effectively accommodate local frustration and packing defects, leading to a sub-extensive  energy correction.

Finite-size effects on the dynamics provide complementary insight.  
Diffusion offers the most direct probe of the elementary rearrangement scale, as it is less sensitive to long-range dynamical facilitation and elastic coupling than the structural relaxation time.  
Figure~\ref{fig1}c shows the temperature dependence of the inverse diffusivity for all system sizes.  
A clear fragile-to-strong crossover (FSC) emerges at $T\lesssim 0.1$ for every $N$.  After fitting the low-temperature and high-temperature range in Figure~\ref{fig1}c with simple activated behavior, we report the ratio of both values, see  inset of Fig.~\ref{fig1}c. 
Even $N=33$ reproduces the qualitative behaviour of the thermodynamic limit, while $N=66$ already yields nearly quantitative agreement.  
The slight non-monotonicity with $N$ is plausible: very small systems suppress cooperative rearrangements, whereas intermediate sizes may occasionally overrepresent rare excitations; only once $N$ exceeds the characteristic cooperative length do these effects balance out, producing bulk-like behaviour.  
This has qualitative similarities to finite-size effects seen in the Kac–Fredrickson–Andersen model~\cite{berthier2012finite}, though a direct correspondence remains open. 

Finally, the choice $N_c = 66$ is supported by structural indicators.  
The participation ratio—which measures the number of particles involved in an IS–IS transition—shows only weak $N$-dependence for $N\ge N_c$~\cite{vogel2004particle} (SM, Sec.~II).  
Taken together, the thermodynamic scaling (fluctuations and mean energies), the near size-independence of diffusion, and the plateau of the participation ratio all identify $N_c = 66$ as the minimal system size (of those, studied in this work)  that simultaneously reproduces bulk behaviour in thermodynamics, dynamics, and structural features while remaining small enough to allow complete PEL sampling.

\subsection{Accessing the PEL: Gaussian regime, depletion, and cutoff}\label{sec:depletion_PEL}

\begin{figure}[b]
    \includegraphics[scale=1]{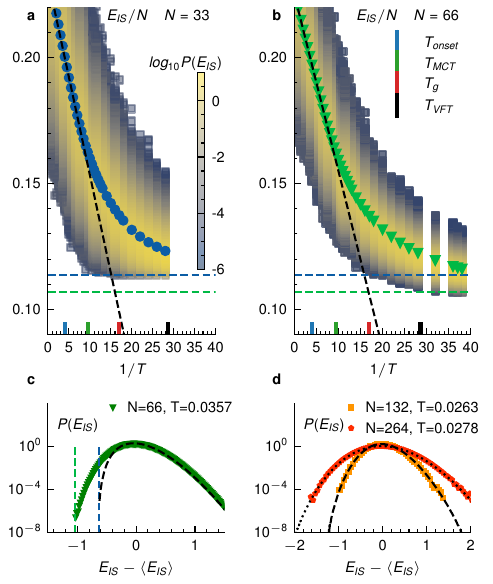}
    \caption{
    The per-particle potential energy of IS with system sizes ({a}) N=33 and ({b}) N=66, respectively. The color bar for each temperature represents the equilibrium probability distribution. 
    { The deviation from the observed Gaussian behavior in the initial supercooled regime is stressed with the dashed line fitted for the temperature range $T\in[0.1429,0.25]$}.
    The horizontal lines represent the lowest energy {of IS} for both system sizes.
    {(c \& d)} The Boltzmann energy distributions, for $N\in[66,264]$  compared with the distribution {  {of subsystems of size}} $N/2$, suggest the subsystem does satisfactorily characterize the larger system. Since the $N=66$ system accesses much lower energies than the $N=33$ system, deviations for this comparison have to occur in the low-energy region (as in {(c)}). 
    }\label{fig2}
\end{figure}

The distribution of IS  energies provides direct insight into how different regions of the potential-energy landscape (PEL) are explored upon cooling.  
Figures~\ref{fig2}(a,b) show the Boltzmann-weighted distributions $P(E_{IS},T)$ for $N=33$ and $N=66$, together with the corresponding averages $\langle E_{IS}\rangle(T)$.  
For temperatures $T>0.1$, the mean IS energy varies linearly with $1/T$, indicating that the sampled part of the landscape is purely Gaussian~\cite{heuer2000density}. The energies in this regime are nearly size independent (SM, Sec.~IV).

Below $T\approx 0.1$, $\langle E_{IS}\rangle(T)$ bends away from linearity and $P(E_{IS},T)$ develops an asymmetric low-energy tail.  
Both features signal the onset of a depletion regime in which the finite number of low-energy states becomes thermodynamically relevant.  
{\rev{The temperature range over which this non-Gaussian regime emerges overlaps, for all $N$, with the fragile-to-strong crossover in the dynamics (Sec.~\ref{sec:thermo_dyn}), indicating a correlation between thermodynamic changes and dynamical behaviour.}}

Upon further cooling, the bottom of the PEL becomes accessible.  
For $N=33$ this occurs near $T\approx 0.06$ (close to the calorimetric $T_g$), whereas for $N=66$ the lowest states are reached at significantly lower temperature (around $T_g/2$).  
Independent statistical arguments (Appendix~\ref{sec:bottom},~\ref{sec:lowspectrum}) {\color{black} strongly suggest} that the global minimum is indeed found.  
Thus, for the first time, the complete IS equilibrium distribution - including the full low-energy spectrum - is available for a non-network-forming system of size $N_c=66$.

Because the minimum IS energy decreases with increasing system size, the temperature at which these lowest states become thermally populated, denoted $T_{\mathrm{cut}}(N)$, must decrease with $N$.  
This trend is observed directly in our data and rationalized in Appendix~\ref{sec:bottom} and SM, Sec.~V.  
Whether the corresponding crossover size $N_c$ diverges strictly only as $T\to 0$ cannot be determined unambiguously.  
A simple two-state model (SM, Sec.~IV.G) and earlier theoretical analyses~\cite{berthier2019zero,jung2025numerical} indicate that the probability of occupying the absolute minimum decreases only logarithmically with $N$, implying that $N_c$ remains finite at any nonzero temperature, while a divergence as $T\to 0$ cannot be excluded.  
In practice, complete sampling of the PEL is feasible for $N_c=66$ over all experimentally relevant temperatures, whereas for larger systems the required temperatures (e.g.\ $T\lesssim 0.02$ for $N=132$) lie far below feasible equilibrium conditions.

\begin{figure*}
    \includegraphics[scale=1]{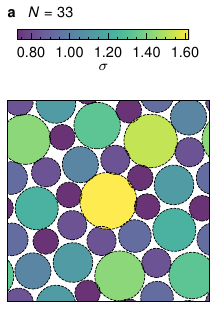} \hspace{-3mm}
    \includegraphics[scale=1]{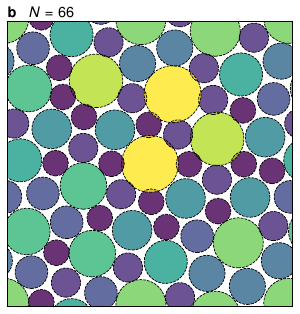}
    \includegraphics[scale=1]{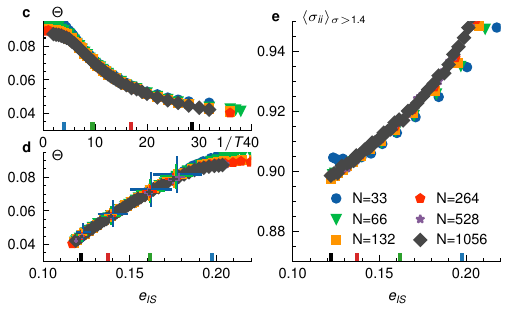}
    \caption{
    {(a \& b)} The lowest energy sample was explored during a wide range of temperatures. For N=33(66), the lowest energy configuration is shown for T=0.0385(0.0270) and T=0.0357(0.0263) with filled and dashed styles, respectively. The color scheme represents the diameter of the particle.  
    {(c)} The disorder depicts a similar temperature dependence as the energy.  
    {(d)} The parametric plot between the structural disorder and the potential energy provides a comprehensive approach to studying the disorder.
    {(e)} The mean diameter for the neighbors of the particle $i$ with diameter $\sigma_i\ge 1.4$ for a range of temperatures across $T_g$. The distinct nature of N=33 suggests that the variety of possible configurations is not large enough for the smallest system. 
    }\label{fig3}
\end{figure*}

To test the superposition hypothesis, Fig.~\ref{fig2}(c,d) compares the measured distributions $P(E_{IS},T)$ with those reconstructed by convolving the distributions of systems of size $N/2$ (see Eq/~\ref{eq:superposition}).  
For $N=33\rightarrow 66$ the agreement is excellent except near the very lowest energies, where different ground-state energies naturally cause deviations.  
For $N=66\rightarrow 132$ and $132\rightarrow 264$ the superposition holds over the \emph{entire} sampled energy range, including the low-energy tail.  
This demonstrates that $N_c=66$ is the smallest system size that simultaneously allows complete access to the low-energy states and exhibits bulk-like PEL behavior. {\rev{The distinct energy cutoffs for $N=33$ and $N=66$ indicate that finite-size effects become relevant at low temperatures below a system-size-dependent threshold; for the present system, $N=66$ provides a reliable description of bulk behaviour for $T \gtrsim T_g/2$ (as discussed in SM Sec.~IV(E–G)).}}

\subsection{Deep Minima and Structural Ordering}
Figure~\ref{fig3}(a,b) shows the lowest inherent structures  obtained for $N=33$ and $N=66$ at two closely spaced temperatures in the regime where the bottom of the potential-energy landscape (PEL) becomes populated.  
For each size, the resulting configurations coincide within numerical precision, confirming that the same ground state is reached repeatedly and that the minimum identified in Appendix~\ref{sec:bottom} (and Sec.~\ref{sec:depletion_PEL}) indeed corresponds to the true PEL bottom.  
This consistency between statistical arguments (Appendix~\ref{sec:lowspectrum}) and equilibrium sampling provides strong evidence that the full low-energy spectrum is resolved.

To quantify structural changes with decreasing temperature, we use the disorder parameter $\Theta$ introduced in Ref.~\cite{tong2018revealing}, which measures local angular deviations from steric arrangements.  
As shown in Fig.~\ref{fig3}(c), $\Theta(T)$ closely follows the temperature dependence of the IS energy and saturates already above $T_g$.  
This behaviour originates from the nearly linear correlation between disorder and IS energy (Fig.~\ref{fig3}(d)), and from the presence of a structural cutoff in the PEL.  
Because $\Theta$ is experimentally accessible, this correlation provides a potential structural fingerprint of the approach to the PEL bottom.

Another distinguishing feature of low-energy configurations is the emergence of strong large–small anticorrelations.  
Figure~\ref{fig3}(e) reports the mean diameter of Voronoi neighbours for particles with $\sigma_i\!\ge\!1.4$.  
Upon cooling, these particles increasingly neighbour smaller ones, allowing efficient packing and access to lower energies.  
This trend is robust for all $N\!\ge\!N_c$, while deviations for $N=33$ reflect the more limited set of accessible packings.  
The observed anticorrelation aligns with the recently discovered ``exotic compositional order’’ in non-additive mixtures~\cite{tong2023emerging} and plays a role analogous to the disappearance of well-defined defects at low energies in network-forming liquids~\cite{roy2019ring}.

Overall, the approach to the PEL bottom is marked by reproducible ground-state configurations, saturation of local disorder, and highly efficient local packing.  
These structural signatures provide the microscopic foundation for the depleted low-energy density of states and reinforce the thermodynamic analysis of the next section.

\subsection{Landscape-Driven Origin of the FSC}
\label{sec:thermo_dyn}
Having fully characterized the depleted low-energy density of states, we now demonstrate that this thermodynamic feature directly dictates the dynamical behavior, including the fragile-to-strong crossover (FSC).
Here we show that this correlation {\rev{is expected if we describe the PEL as a trap model. The following assumptions enter this analysis which have been shown to hold for a binary LJ-system~\cite{doliwa2003finite,heuer2008exploring,rehwald2010coupled} : (1) For low energies, the Boltzmann distribution of IS energies is very similar to the distribution of metabasin energies. (2) The diffusivity is proportional to the average escape rate from the metabasins, visited at a given temperature $T$(=$1/k_B\beta$). (3) The average escape rate is proportional to $\exp(-\beta (V_0-E))$. Here $E$ is defined on the length scale of one cooperatively rearranging region, characterizing the transition between subsequent metabasins and $V_0$ the a priori unknown effective energy level of saddle points. }}

With these assumptions we may write 
\begin{equation}
    D(\beta) \propto \langle \Gamma \rangle(\beta)
    = \sum_i P(E_{i,IS},\beta)\,
      \exp[-\beta (V_0 - E_{i,IS})].
\end{equation}
Taking the derivative of the logarithm yields the apparent activation energy
\begin{equation}
  E_{\mathrm{app}}(\beta)
  = -\frac{d}{d\beta}\ln \langle \Gamma \rangle(\beta).
\end{equation}
As shown in Appendix~\ref{sec:analy}, within a trap-like physical picture the apparent activation energy can be directly related to the average IS-energy via
\begin{equation}
    E_{\mathrm{app}}(\beta)
    = V_0 - \langle E_{IS} \rangle(\beta).
    \label{eq:simple_relation}
\end{equation}
It is valid for \emph{any} distribution $G(E_{IS})$ of IS energies.
Thus, any change of the {\it thermodynamic} observable $\langle E_{IS} \rangle(\beta)$ caused by the depletion regime directly implies a corresponding change in the apparent activation energy, i.e.\ in the {\it dynamics}, giving rise to a weaker temperature dependence and ultimately to a FSC.

\begin{figure}[] 
    \includegraphics[width=\linewidth]{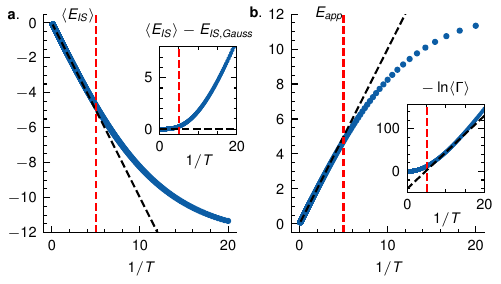}
    \caption{
    {(a)} The average IS energy calculated numerically for a binomial distribution of states ($M=150$) shows a linear regime at higher temperatures, and deviation from the Gaussian-PEL at the low temperature (inset). 
    {(b)} The apparent activation energy from hopping rate shows a deviation from the high-temperature behavior, reflecting the relation given by Eq.~\ref{eq:simple_relation}. 
    (b, inset) A logarithmic representation of the   hopping rate, with the tangent used to determine the activation energy at $1/T=10$.
    The FSC is marked with vertical red line.}
    \label{fig:binomial}
\end{figure}

To make this mechanism explicit, Fig.~\ref{fig:binomial} analyses a binomial density of states,
\begin{equation}
 G(E_{IS}) = \binom{M}{i} 2^{-M},
\end{equation}
M is the number of energy states of system, and $i$ is index of those states. 
The binomial density of states is Gaussian at high energies (by the central limit theorem) but possesses a sharp, known cutoff at its lowest energy $E_{\min} = -\sigma_K \sqrt{M}$ (see Appendix~\ref{sec:binom} for details). This simple model allows us to track \emph{exactly} how the depletion regime influences both thermodynamics and dynamics.

Figure~\ref{fig:binomial}(a) shows the average IS energy $\langle E_{IS}\rangle(\beta)$ for such a landscape. For high temperatures the behaviour is strictly Gaussian and linear in inverse temperature. As temperature is lowered, the mean energy bends away from this linear extrapolation once the finite number of low-energy states becomes relevant; this deviation from the Gaussian reference with identical variance is highlighted in the inset. 
Via Eq.~\eqref{eq:simple_relation}, the consequences for the dynamics follow immediately. The average hopping rate leads to the apparent activation energy shown in Fig.~\ref{fig:binomial}(b), which is essentially the mirror image of $\langle E_{IS}\rangle(\beta)$. 
{\color{black} At low $T$, the hopping rate as a function of $1/T$ gradually approaches Arrhenius behaviour, while the apparent activation energy continues to vary due to the finite cutoff of the density of states. This illustrates that in this fully controlled setting (1) the FSC is the dynamical manifestation of the thermodynamic depletion of low-energy states and (2) a slower increase of the apparent activation energy may easily appear as a simple Arrhenius behavior (in particular if additional noise were added).

As shown in Appendix E, a small–$\beta$ expansion gives
\begin{align}
   \langle E_{IS}\rangle(\beta)
   \approx -\beta\sigma_K^2
   + \frac{\beta^3\sigma_K^4}{3M}.
\end{align}
This has the interesting consequence that the onset of depletion can be systematically shifted by varying the parameter $M$, which controls the effective number of available low-energy states. For large $M$, the Gaussian regime extends to lower energies and depletion becomes relevant only at very low temperatures. In the limit $M \to \infty$, the deviation from Gaussian behaviour is shifted towards $T \to 0$, and no crossover would be observed within an experimentally accessible temperature window. If a glass transition temperature $T_g$ is defined via a dynamical criterion (as introduced above), decreasing $M$ (starting from very large $M$) shifts the onset of depletion relative from below $T_g$  to finally above $T_g$. Correspondingly, the FSC may occur above, close to, or below $T_g$. In the latter case, the system would appear fragile over the entire experimentally accessible range, even though the PEL still possesses a lower bound and an associated depletion regime at sufficiently low temperatures. 
}

\begin{figure}[] 
    \includegraphics[width=0.9\linewidth]{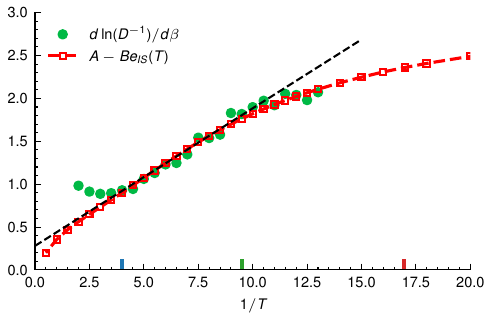}
    \caption{
    The apparent activation energy (with $A= 5.69$ and $B= 24.29$) and the IS energy, from the simulation, show comparable temperature dependence. For $T \gtrsim 0.1$, both vary linearly, while deviations at lower temperatures mark the onset the FSC. This correspondence supports a common thermodynamic origin, consistent with the trap-model framework.}
    \label{fig:connection}
\end{figure}

{\color{black}
We now test whether the relation derived above,
Eq.~\eqref{eq:simple_relation},
is consistent with the numerical simulation results.
Figure~\ref{fig:connection} compares the temperature dependence of the
average IS energy per particle,
$e(T)=\langle E_{IS}\rangle/N$, with the apparent activation energy
$E_{\mathrm{app}}(T)$ obtained from the temperature dependence of the diffusivity.

As expected theoretically, the thermodynamic quantity $e(T)$
is approximately linear in $1/T$ in the Gaussian regime of the PEL
($1/T \lesssim 10$).
Upon cooling, $e(T)$ gradually deviates from this linear behaviour and
approaches saturation as the finite number of low-energy amorphous states
becomes relevant.
This deviation marks the thermodynamic onset of the depletion regime.

According to Eq.~\eqref{eq:simple_relation},
$E_{\mathrm{app}}(T)$ should exhibit the same temperature dependence as
$\langle E_{IS}\rangle(T)$ up to a linear transformation.
To test this prediction, we introduce the mapping
\begin{equation}
E_{\mathrm{app}}(T) \simeq A - B\, e(T),
\label{eq:fit_Eapp_e}
\end{equation}
where $A$ and $B$ are determined from the high-temperature data in the
Gaussian regime.
Figure~\ref{fig:connection} shows that after this transformation the two
curves nearly coincide over the full investigated temperature range
$4 \le 1/T \le 13$.
The agreement therefore extends beyond the Gaussian regime into the
low-temperature region where depletion effects become relevant.

The gradual saturation of $e(T)$ is accompanied by a corresponding
slowdown in the increase of $E_{\mathrm{app}}(T)$, indicating the onset
of the fragile-to-strong crossover (FSC).
This crossover is continuous and does
not correspond to a sharply defined temperature.
For $e(T)$ the crossover can be quantified directly from the deviation
from Gaussian behaviour, while for $E_{\mathrm{app}}(T)$ the present data
capture the initial approach towards the low-temperature Arrhenius regime.

Within the trap-based interpretation introduced above,
relaxation corresponds to activated transitions between metabasins. If we denote the characteristic trap size as $N_{\mathrm{trap}}$, then the energy on this scale, relevant for the trap model, is given by  
\begin{equation}
E_{\mathrm{trap}}(T) = N_{\mathrm{trap}}\, e(T).
\end{equation}
Comparison of Eqs.~\eqref{eq:simple_relation} and
\eqref{eq:fit_Eapp_e} yields
\begin{equation}
A \approx V_0, \qquad B \approx N_{\mathrm{trap}},
\end{equation}
allowing us to estimate a characteristic trap size
$N_{\mathrm{trap}} \approx 20$–$30$ particles.

This quantity should not be interpreted as a geometric subsystem size
in the sense of the thermodynamic crossover size $N_c$, which denotes the
smallest system size that reproduces bulk PEL statistics.
Instead, $N_{\mathrm{trap}}$ is a dynamical quantity describing the number
of degrees of freedom that collectively determine the depth of a metabasin
and thus the effective activation barrier.

Consistently, $N_{\mathrm{trap}}$ exceeds the participation ratio obtained
from individual IS–IS transitions (SM Sec.~II), which typically involves
about ten particles.
The participation ratio measures the active core of a local rearrangement,
whereas $N_{\mathrm{trap}}$ characterizes the larger collective entity
that stabilizes the metabasin, including the surrounding elastic degrees
of freedom.
Similar relations between local rearrangements and metabasin-level
collectivity have been observed previously for Lennard-Jones systems
\cite{doliwa2003finite,vogel2004particle}.

Taken together, the numerical results show that the temperature dependence
of the apparent activation energy closely follows that of the IS
energy, as predicted by Eq.~\eqref{eq:simple_relation}.
The progressive depletion of low-energy states therefore provides a
consistent thermodynamic explanation for the gradual FSC observed in the
dynamics.
In this picture, the FSC does not require an additional dynamical mechanism,
but reflects the changing statistical properties of the low-energy region
of the potential-energy landscape.}

\begin{figure*}[]
    \includegraphics[width=\textwidth]{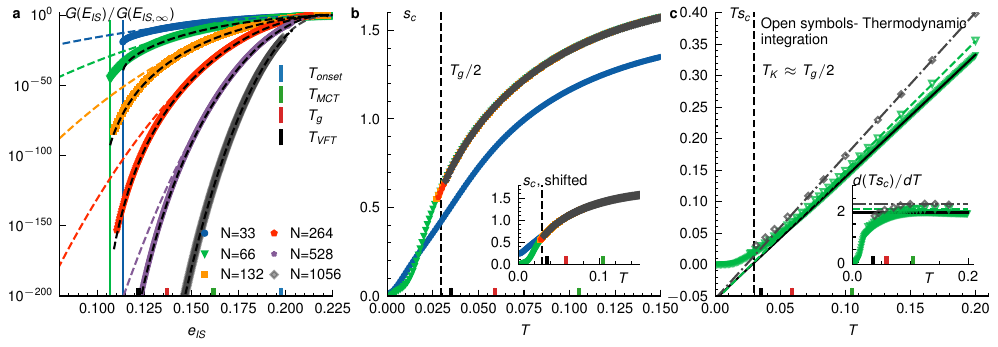}
    \caption{
    {(a)} The energy distribution of IS for various system sizes across the ultra-stable limit. 
    The dashed-colored lines show a Gaussian fit from the higher energy supercooled regime, revealing a faster depletion of states than predicted for a pure Gaussian.
    The dashed-black line depicts the distribution constructed from the respective half-sized subsystem. 
    The vertical-colored lines represent the lowest energy for IS for N=33 and 66. The colored markings on the $x$-axis indicate the specific temperatures and the corresponding potential energies used in the analysis. 
    {(b)} For $N=33, 66$, the configurational entropy is calculated from the complete IS energy distribution, including the information of the {{basin's volumes} from} the Hessian, i.e. the matrix of second derivaties at the minimum. For $N\geq 132$, the entropy curves are shifted to agree with N=66 at $T_{MCT}=0.105$. 
    {The inset compares the `shifted' entropy for $N=33$ to $N=66$ at $T_{MCT}$ and its range of (nearly) perfect agreement. The vertical dashed line represents $T_g/2$.}
   {(c)}  The comparison of the linearised configurational entropy ($Ts_c$) for N=66 (partition sum) and for N=66 and N=1056 (thermodynamic integration). The fit lines represent the conventional `linear fit' to estimate Kauzmann's temperature($T_K$). The inset shows the derivative of $T s_c$ to highlight the loss in linearity at low temperatures.}\label{fig4}
\end{figure*}

\subsection{Configurational Entropy of the Depleted PEL}

The thermodynamics of the system follows from the volume-weighted density of inherent structures,
\begin{equation}
    G(E_{IS}) = \sum_i Y_i\,\delta(E_{IS}-E_{i,IS}),
\end{equation}
which determines the equilibrium probability
$P(E_{IS},T)\propto G(E_{IS})\,e^{-\beta E_{IS}}$.  
{\rev{The basin-volume factors $Y_i$, configurations that relax to the same IS, are obtained within the harmonic approximation from the Hessian eigenvalues (second derivatives of the potential energy)~\cite{sciortino2005potential}.}}
We determine $G(E_{IS})$ via reweighting of sampled IS energies~\cite{saksaengwijit2004origin} ({\rev{details in }}Appendix~\ref{sec:entropypart}, SM Sec.~IV).

Figure~\ref{fig4}(a) shows density $G(E_{IS})$ of IS energies for several system sizes.  
At high energies the density follow Gaussian-distribution, while at low energies it exhibits a pronounced depletion, consistent with the structural signatures discussed in Sec.~V and the deviations in $P(E_{IS},T)$ shown in Fig.~\ref{fig2}. 
The superposition hypothesis holds for all $N\!\ge\!66$ across the entire energy range; deviations in the $N=33\rightarrow 66$ comparison arise only near the cutoff because the minima differ.

The configurational entropy follows from the Shannon expression,
\begin{equation}
S_c(T)=-\sum_i P(E_{i,IS},T)\ln P(E_{i,IS},T),
\end{equation}
which can be rewritten as
\begin{equation}
S_c(T)= -\langle \ln Y_i\rangle
        + \beta\langle E_{IS}\rangle
        + \ln\!\Big(\sum_i Y_i e^{-\beta E_{i,IS}}\Big),
\end{equation}
further discussed in Appendix~\ref{sec:entropypart}.  
The final term requires the \emph{absolute} number of inherent structures and is therefore only available for $N=33$ and $N$(=$N_c$)= $66$, for which the complete low-energy spectrum—including the ground state—is known.  
This permits the first full determination of $S_c(T)$ for a non-network-former without reference to liquid-state entropy.

Figure~\ref{fig4}(b) displays the resulting entropy curves.  
For $N=66$, $S_c(T)$ transitions from positive to negative curvature when entering the depletion regime, reflecting the geometric constraint imposed by the structural cutoff.  
Comparison with thermodynamic integration (Fig.~\ref{fig4}(c)) shows excellent agreement in the overlapping temperature window, demonstrating very small finite-size effects.  
The breakdown of linearity in $T s_c(T)$ at low temperatures is a direct consequence of the depleted low-energy density and implies that, within the finite-size PEL description, the entropy vanishes only as $T\to 0$, consistent with the avoidance of a finite-temperature entropy crisis~\cite{kauzmann1948nature,angell2008glass} and with recent numerical analyses~\cite{berthier2019zero}.  
For $T<0.01$, the entropy reflects the discrete spacing of the low-energy spectrum (Appendix~\ref{sec:lowspectrum}).

A notable observation emerges when comparing $N=33$ and $N=66$.  
Although their average energies and diffusivities are nearly identical across all temperatures, the absolute configurational entropy differs significantly:  
$N=66$ exhibits roughly $\exp[(s_c(66)-s_c(33))\cdot 66]\approx 10^7$ more distinct inherent structures.  
This follows naturally from factorisation arguments and can be rationalized by an approximate $\ln(N)/N$ finite-size contribution to the absolute entropy (Appendix~\ref{sec:entropypart}, SM Sec.~IV.F).

Taken together, these thermodynamic results show that the depleted low-energy density, structurally rooted in Sec.~(B\& C), governs both the shape of the configurational entropy and its approach to zero in the ideal-glass limit.  
This thermodynamic behaviour directly underlies the fragile-to-strong crossover discussed in the following section. \\\\ 
\section{Conclusion}
{\color{black}
By resolving the potential-energy landscape (PEL) down to the amorphous ground state of a finite non-network-forming model system, the present work establishes a microscopic picture of thermodynamic and dynamical behaviour in the deeply supercooled regime. Identifying a characteristic system size $N_c$ that is small enough for exhaustive sampling yet large enough to reproduce bulk PEL statistics above $T_g/2$ enables reconstruction of the full density of IS energies for that size, including the lowest-energy configurations and their excitations. This allows the configurational entropy to be determined directly from the landscape over the full temperature range without relying on liquid-state thermodynamic integration.

A central thermodynamic result is the pronounced depletion of low-energy states relative to the Gaussian regime of the PEL, which determines the curvature of the configurational entropy at low temperatures. Within the IS framework applied to the finite systems studied here, the configurational entropy approaches zero only as $T \to 0$. This behaviour reflects the existence of a lower bound of the PEL; whether it persists in the thermodynamic limit remains an open question, since the Kauzmann temperature $T_K$ is defined only for macroscopic systems.

The same depletion regime is also reflected in the dynamics. Numerically, the apparent activation energy closely follows the temperature dependence of the mean IS energy over the entire investigated range. As temperature decreases, the progressive depletion of low-energy states is accompanied by a gradual crossover from fragile to stronger temperature dependence of the diffusivity, without implying a sharply defined transition temperature. This correlation is consistent with a trap-based interpretation of the PEL, in which the apparent activation energy is directly linked to the mean IS energy.

From the linear relation between apparent activation energy and per-particle IS energy, we estimate a characteristic dynamical size $N_{\mathrm{trap}}$ of roughly two dozen particles. This quantity reflects the number of degrees of freedom that collectively determine the depth of a metabasin and the associated activation barrier and is conceptually distinct from the thermodynamic sampling size $N_c$. Consistently, $N_{\mathrm{trap}}$ exceeds the participation ratio obtained for individual IS–IS transitions (SM Sec.~II), indicating that metabasin stability involves both the locally rearranging core and its surrounding elastic environment. Such a separation of scales has also been observed in Lennard-Jones systems \cite{doliwa2003finite,vogel2004particle} and may provide a physically motivated basis for coarse-grained descriptions of activated dynamics such as elasto-plastic models \cite{rehwald2010coupled,ozawa2023elasticity}.

The resolved low-energy structures further reveal pronounced structural signatures, including strong anticorrelations between large and small particles, efficient local packing, and saturation of local disorder. These features appear already in the smallest systems and are consistent with recent observations of compositional ordering tendencies in non-additive mixtures \cite{tong2023emerging}, representing an analogue—here for non-network formers—of defect minimization mechanisms known from network-forming liquids \cite{roy2019ring}. Such motifs are likely related to the excellent glass-forming ability of the present system.

More generally, the results illustrate how the existence of a lower bound of the PEL can influence both thermodynamics and dynamics at low temperatures. For any finite system, a depletion of low-energy states relative to the Gaussian regime must occur sufficiently close to the bottom of the landscape. If relaxation proceeds as envisaged by the trap model this depletion is expected to induce a gradual crossover towards Arrhenius behaviour as the bottom of the PEL is approached. Depending on how rapidly the density of accessible amorphous states decreases, the onset of depletion may lie either above or below $T_g$. Consequently, a fragile-to-strong crossover may be observable within the experimentally accessible temperature range, but it may also occur only at lower temperatures, in which case the system appears fragile throughout the relevant regime. Experimental studies of several glass formers indeed show indications of Arrhenius-like behaviour already above $T_g$ (SM Sec.~VI), consistent with this general scenario. Another well-known example is silica with a FSC far above $T_g$.

An important open question concerns how these thermodynamic and dynamical signatures evolve with increasing system size at very low temperatures. Additional complexity arises from the growth of cooperative length scales and possible changes in the effective number of relevant configurational degrees of freedom. How the depletion regime and its dynamical consequences develop in the thermodynamic limit therefore remains an interesting topic for future work.
 }

\section*{Supplementary material} 
The Supplementary Material provides additional information on the analyses and computational procedures underlying the results and conclusions presented in this work. Specifically, it includes details on:
(i) sampling in perfect equilibrium;
(ii) identification of particles participating in the relaxation process;
(iii) characterization of the dynamics;
(iv) finite-size effects; and
(v) the absence of crystallization effects; and 
(vi) FSC Signatures in experimental glass formers. \\\\
{\bf Acknowledgment} We greatly acknowledge the usage of the HPC-system PALMA2 at the University of Münster, Germany. \\\\
{\bf Competing interests} The authors declare no competing interests. \\\\
{\bf Data Availability}
The data and the code can be found under \href{https://github.com/anshuldsp/depleting-states-ideal-glass}{https://github.com/anshuldsp/depleting-states-ideal-glass}.

\section*{Appendix}
\appendix
\section{The temperature associated with the identification of the bottom of the PEL}\label{sec:bottom}
\begin{figure}[h]
    \includegraphics[scale=1.00]{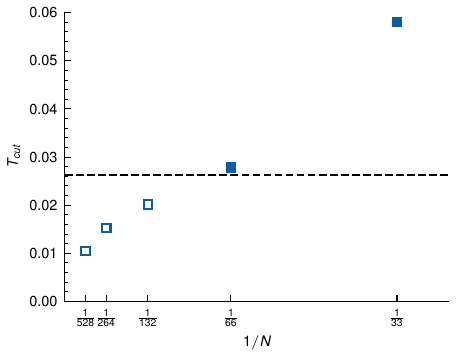}
    \caption{
    The $T_{cut}$ vs. $1/N$ for a large range of system sizes. The horizontal line represents the lowest temperature explored in equilibrium with the swap Monte-Carlo simulation. 
    To estimate $T_{cut}$ for $N\ge132$(open symbols), the complete IS density has been estimated via the factorization hypothesis.}\label{SI:fig8}
\end{figure}

To determine the temperature at which the lowest IS energy becomes thermally accessible, we introduce a well–defined cutoff temperature $T_{\mathrm{cut}}$.  
Rather than relying on whether the lowest state happens to be sampled in a finite trajectory, we define $T_{\mathrm{cut}}$ through a fixed probability criterion,
\begin{equation}
P(E_{\mathrm{lowest}},T_{\mathrm{cut}}) = p_{\mathrm{cut}},
\end{equation}

with $p_{\mathrm{cut}} = 10^{-6}$.  
This value reflects the empirical probability with which the lowest IS was sampled for $N=33$ and $N=66$ at the onset of consistent ground–state identification.  
The resulting $T_{\mathrm{cut}}$ values closely match the temperatures at which the ground state is actually observed in simulations (Fig.~\ref{SI:fig8}), confirming the robustness of this operational definition.

To predict $T_{\mathrm{cut}}$ for larger systems, we invoke the superposition (factorization) hypothesis: the energy distribution of a system of size $2N$ is the convolution of two statistically independent subsystems of size $N$.  
Using the fully sampled density of states for $N=66$, we can therefore construct the predicted distributions for $N=132$ and beyond and evaluate the corresponding $T_{\mathrm{cut}}$.  
As shown in Fig.~\ref{SI:fig8}, $T_{\mathrm{cut}}$ decreases rapidly with increasing system size.

This trend has a simple statistical origin.  
Even if the lowest IS energy were the same for $N$ and $2N$, the probability of occupying that state at a given temperature decreases exponentially with the number of independent subsystems.  
Consequently, identifying the absolute minimum requires progressively lower temperatures as $N$ grows.  
The difference between $T_{\mathrm{cut}}$ for $N=33$ and $N=66$ therefore reflects not only the different minimum energies but also this generic statistical suppression.

Under the idealized assumption that the factorization hypothesis holds down to the very bottom of the landscape, one would expect that a macroscopic system requires $T_{\mathrm{cut}}\to 0$ in order to sample its global minimum with probability of order unity.  
This vanishing of $T_{\mathrm{cut}}$ is thus a statistical effect rather than a signature of singular thermodynamics.

For the present study, the relevant observation is that the smallest system size reproducing bulk thermodynamic and dynamical behaviour, $N_c = 66$, also allows full sampling of the entire PEL down to its true minimum within accessible temperatures.  
For larger $N$, the temperatures required to identify the global minimum ($T_{\mathrm{cut}}\lesssim 0.02$ for $N=132$) lie far below feasible equilibrium conditions, but this does not affect any experimentally relevant temperature regime or the interpretation of the PEL for $T \gtrsim T_g/2$.

\section{Low energy spectrum near the cutoff}\label{sec:lowspectrum}

\begin{figure}[h]
    \includegraphics[scale=1.0]{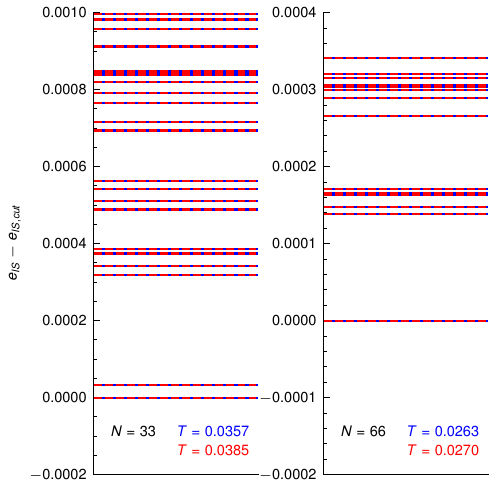}  
    \caption{{{PEL near cutoff: }} 
    The per-particle energy of the IS highlights the low energy spectrum near the cutoff. The energies are shifted with the bottom to show the same energy range. For N=33 and N=66, within the relevant Boltzmann weight, the 28 and 16 energy levels are the same across the lowest two temperatures, respectively.}\label{SI:fig9}
\end{figure}

Figure~\ref{SI:fig9} shows the resolved low-energy spectrum for $N=33$ and $N=66$.  
To place both systems on a comparable scale, the IS energies are shifted such that the respective lowest values satisfy  
$E_{\mathrm{lowest}}/N = 0.113652322808157$ (for $N=33$) and  
$E_{\mathrm{lowest}}/N = 0.106756393732118$ (for $N=66$).  
Two levels $i$ and $j$ are considered identical if their per-particle energy difference satisfies $\frac{|E_{i,IS}-E_{j,IS}|}{N} \le 10^{-14}$,
ensuring numerical uniqueness of each resolved state.

Panels~\ref{SI:fig9} display the IS energies sampled at the two lowest temperatures investigated for each system size  
($T=0.0357$ and $0.0385$ for $N=33$;  
 $T=0.0263$ and $0.0270$ for $N=66$).  
We plot all inherent structures within an energy window
$E_{i,IS} - E_{\mathrm{lowest,IS}} \le T$, 
which identifies the subset of states with non-negligible Boltzmann weight.  
Using this criterion, we find 28 such states for $N=33$ and 16 for $N=66$.

Crucially, \emph{every} low-energy state identified at one temperature is also observed at the other.  
This confirms that our sampling is sufficiently exhaustive to reconstruct the complete low-energy spectrum for both system sizes. In particular, we can thus identify the true ground state and all excitations that remain thermally relevant down to the lowest accessible temperatures.

For $N=66$, the excitation gap between the ground state and the first excited IS is approximately $0.01$.  
Consequently, a clear separation of the ideal glass from the excited states is possible only for temperatures below $T_g/6$, while at higher temperatures the discrete structure of the spectrum becomes thermodynamically irrelevant.

\section{Configurational entropy: partition-sum formulation}
\label{sec:entropypart}

\begin{figure*}[]
    \centering
    \includegraphics[width=\textwidth]{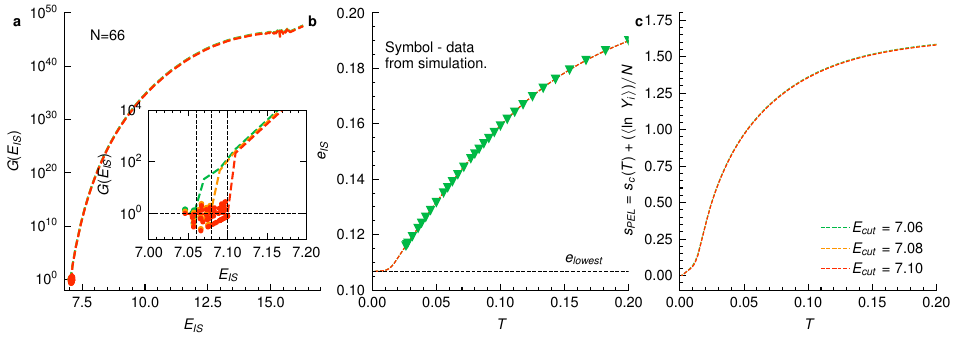}
    \caption{{Thermodynamics of PEL ($N=66$): }
    {(a)} Volume-weighted density of states $G(E_{IS})$ for different choices of the cutoff $E_{\mathrm{cut}}$.
    {(b)} Average IS energy computed from $G(E_{IS})$ and from direct simulations.
    {(c)} $s_c(T) + \langle \ln Y_i\rangle/N$, showing negligible dependence on $E_{\mathrm{cut}}$.}
    \label{SI:fig10}
\end{figure*}

\subsection{Decomposition of the total entropy}

In the canonical ensemble, the total entropy may be written as
\begin{equation}
   S(T) = -\int dx \, P(x,T)\,\ln[P(x,T)u],
\end{equation}
where $u$ denotes a phase–space measure.  
Decomposing phase space into basins $\Omega_i$ of the inherent structures, this becomes
\begin{equation}
   S = -\sum_i \int_{\Omega_i} dx \, P(x,i,T)\,\ln[P(x,i,T)u].
\end{equation}

Introducing the conditional probability $q_i(x)$ within basin $i$ and the probability
$P(E_{i,IS},T)$ to occupy IS $i$, we obtain
\begin{equation}
   S = -\sum_i P(E_{i,IS},T)\!
   \int_{\Omega_i} dx\, q_i(x) 
        \ln\!\bigl[P(E_{i,IS},T)q_i(x)u\bigr].
\end{equation}
This expression separates naturally into
\begin{align}
   S &= -\sum_i P(E_{i,IS},T)\,\ln P(E_{i,IS},T) \nonumber\\
     &\quad -\sum_i P(E_{i,IS},T)
      \int_{\Omega_i} dx\, q_i(x)\ln[q_i(x)u],
\end{align}
which we identify as
\begin{equation}
   S = S_c(T) + S_{\mathrm{vib}}(T).
\end{equation}

The first term,
\begin{equation}
   S_c(T) = -\sum_i P(E_{i,IS},T)\,\ln P(E_{i,IS},T),
   \label{eq:Shannon2}
\end{equation}
is the configurational entropy (Shannon entropy of basin occupations).  
The second term, $S_{\mathrm{vib}}$, represents the vibrational entropy within basins.

\subsection{Harmonic description of basin volumes}~\label{sec:basin_volume}
Both quantities require a characterization of the basin partition sum
\begin{equation}
   z_i(T) = \int_{\Omega_i} dx\, u\, e^{-\beta(E(x)-E_i)}.
\end{equation}
At sufficiently low $T$ the basins are well approximated as harmonic wells around each minimum ($\lambda_{i,\nu}$ are the eigenvalues of the Hessian),
\begin{align}
   z_{i,\mathrm{harm}}(T)
     &= \prod_{\nu=1}^{Nd-d}
        \sqrt{\frac{2\pi}{\beta\lambda_{i,\nu}}}
      = T^{\frac{Nd-d}{2}}\,Y_i,
\end{align}
with $Y_i$ the effective basin volume in harmonic approximation,

\begin{equation}\label{eq:basin_volumn}
   Y_i = \prod_{\nu=1}^{Nd-d} 
              \sqrt{\frac{2\pi}{\lambda_{i,\nu}}}
        = 2\pi^{\frac{Nd-d}{2}}
          \exp\!\left[-\tfrac12\sum_{\nu}\ln \lambda_{i,\nu}\right].
\end{equation}

\subsection{Configurational entropy from the volume-weighted density of states}
The equilibrium probability of IS $i$ is
\begin{equation}
   P(E_{i,IS},T)
   = \frac{Y_i e^{-\beta E_{i,IS}}}{\mathcal{Z}},
   \label{eq:PofE}
\end{equation}
with partition sum
\begin{equation}
   \mathcal{Z}=\sum_i Y_i e^{-\beta E_{i,IS}}.
\end{equation}
Correspondingly, the volume-weighted IS density entering the main text is
\begin{equation}
   G(E_{IS})=\sum_i Y_i\,\delta(E_{IS}-E_{i,IS}).
\end{equation}
(Anharmonic corrections to $Y_i$ are discussed in SM Sec.~X.)

Inserting Eq.~\eqref{eq:PofE} into Eq.~\eqref{eq:Shannon2} yields
\begin{equation}
   S_c(T) = -\langle \ln Y_i\rangle
            + \beta\langle E_{IS}\rangle
            + \ln\mathcal{Z},
\end{equation}
where $\langle \ln Y_i\rangle = \sum_i P_i \ln Y_i$ and  
$\langle E_{IS}\rangle = \sum_i E_i P_i$.  
Importantly, $S_c$ is invariant under global rescaling $Y_i\to\lambda Y_i$.

\subsection{Evaluation of the partition sum}
To evaluate $\mathcal{Z}$ over the entire energy range, we combine:  
(i) the discrete set of explicitly enumerated low-energy IS  
(for $N=33$ and $N=66$), and  
(ii) a coarse-grained, continuum representation for higher energies.

The crossover between discrete and continuous regimes occurs at a chosen cutoff energy $E_{\mathrm{cut}}$.  
Above this energy, we introduce energy bins  as 
$E_{j,\mathrm{bin}}
   = E_{\mathrm{cut}} + (j+\tfrac12)\Delta E $
(with $\Delta E=0.02$) and define the binned density
$ G_{j,\mathrm{bin}}
   = \sum_{i: E_{i,IS}\in[E_{j,\mathrm{bin}}\pm\Delta E/2]}\! Y_i.
$
Up to an overall constant these quantities follow from the reweighting procedure (SM Sec.~II), and for $N=33$ and $N=66$ the absolute normalization $G_{1,\mathrm{bin}} $ is fixed by the exact enumeration of all IS in the first bin above $E_{\mathrm{cut}}$.

The combined density of states is therefore
\begin{align}
   G(E_{IS}) &= \sum_{E_{i,IS}\le E_{\mathrm{cut}}}
                  Y_i\,\delta(E_{IS}-E_{i,IS}) \nonumber\\
             &\quad + \sum_j G_{j,\mathrm{bin}}
                     \,\delta(E_{IS}-E_{j,\mathrm{bin}}),
\end{align}
and the partition sum becomes
\begin{equation}
   \mathcal{Z}
   = \int dE\,G(E_{IS})\,e^{-\beta E_{IS}}.
\end{equation}

The discretization of the continuum part introduces a negligible error provided the energy difference between $E_{\mathrm{cut}}$ and the lowest IS energy significantly exceeds $\Delta E$.  
This is demonstrated explicitly in Fig.~\ref{SI:fig10} for $N=66$:  
the average IS energy computed from $G(E)$ agrees perfectly with the direct simulation average (panel b), and  
$s_c(T)+\langle\ln Y_i\rangle/N  
 = \beta\langle E_{IS}\rangle+\ln\mathcal{Z}$  
is essentially independent of the choice of $E_{\mathrm{cut}}$ (panel c).

\subsection{Systems larger than $N_c$}

For $N>66$, the bottom of the PEL is not identified within accessible temperatures.  
As a result, the absolute normalization of $G(E_{IS})$ is unknown, implying that $\mathcal{Z}$ is only determined up to a multiplicative constant.  
This affects the entropy by an additive constant (shifting $\ln\mathcal{Z}$) but leaves all energy-derived quantities, including $\langle E_{IS}\rangle$ and the temperature dependence of $s_c(T)$, unchanged.  
This is consistent with Fig.~\ref{SI:fig10} and with the main-text argument that only $N=N_c$ permits an absolute determination of the configurational entropy over the full temperature range.

\section{Relationship between average IS energy and apparent activation energy}
\label{sec:analy}

In this appendix, we provide a short derivation of the identity
$
E_{\mathrm{app}}(\beta) = V_0 - \langle E_{IS} \rangle(\beta),
$
used in Sec.~\ref{sec:thermo_dyn} (Eq.~\ref{eq:simple_relation}). 
The starting point is the normalized Boltzmann distribution of inherent–structure energies,
\begin{align}
    P(E_{i,IS},\beta)
    = \frac{G(E_{i,IS})\,\exp[-\beta E_{i,IS}]}%
           {\mathcal{Z}},
\end{align}
where the partition sum is
$\mathcal{Z}=\sum_j G(E_{j,IS})\,\exp[-\beta E_{j,IS}]$. For an activated escape process in which the rate from IS `$i$' is
\begin{equation}
   \Gamma_i(\beta)=\exp[-\beta(V_0 - E_{i,IS})],
\end{equation}
the thermally averaged hopping rate reads
\begin{align}
   \langle \Gamma\rangle(\beta)
   &= \sum_i P(E_{i,IS},\beta)\,\Gamma_i(\beta) \nonumber\\
   &= \frac{\sum_i G(E_i)\exp[-\beta E_i]\,
                       \exp[-\beta(V_0-E_i)]}{\mathcal{Z}}  \nonumber\\
   &= \frac{e^{-\beta V_0}\sum_i G(E_i)}{\mathcal{Z}},
\end{align}
where we have suppressed the IS index on $E_i$ for clarity.

The apparent activation energy is defined as: 
$E_{\mathrm{app}}(\beta)
   = -\frac{d}{d\beta}\ln \langle \Gamma\rangle(\beta)$. 
Using this expression,
\begin{align}
   \ln \langle \Gamma\rangle(\beta)
   &= -\beta V_0 + \ln\Bigl(\sum_i G(E_i)\Bigr) - \ln \mathcal{Z},
\end{align}
where the second term is $\beta$–independent.  
Taking the derivative with respect to $\beta$ gives
\begin{align}
   E_{\mathrm{app}}(\beta)
   &= V_0 + \frac{d}{d\beta}\ln\mathcal{Z}.
\end{align}
Evaluating the derivative of the partition sum,
\begin{equation}
       \frac{d}{d\beta}\ln \mathcal{Z}
   = -\,\frac{\sum_i E_i G(E_i)\exp[-\beta E_i]}%
            {\sum_j G(E_j)\exp[-\beta E_j]}
   = -\langle E_{IS}\rangle(\beta),
\end{equation}
with which one finally recovers within the trap-like activated-dynamics picture the exact identity
$E_{\mathrm{app}}(\beta) = V_0 - \langle E_{IS}\rangle(\beta)$ (see Sec.~\ref{sec:thermo_dyn}),
which holds for \emph{any} distribution $G(E_{IS})$.

\section{Analytical investigation of a binomial density of states}
\label{sec:binom}

To illustrate analytically how a finite low–energy cutoff modifies the thermodynamics and leads to deviations from Gaussian behaviour, we consider a simple discrete model in which the density of IS energies is binomially distributed.  
Because the energies are indexed by an integer $i$, the model allows an explicit implementation of the lower bound.

The density of states is
\begin{align}
   G(E_{i,IS}) = \binom{M}{i}\left(\tfrac{1}{2}\right)^M,
\end{align}
and the corresponding inherent–structure energies are chosen as
\begin{align}
   E_{i,IS} = \left(i-\tfrac{M}{2}\right)\,\sigma_K\,\frac{2}{\sqrt{M}},
\end{align}
yielding a distribution of mean $0$ and variance $\sigma_K^2$.  
The lowest accessible energy is therefore
\begin{align}
   E_{0,IS} = -\sigma_K\sqrt{M}.
\end{align}

To obtain the average IS energy,
$\langle E_{IS}\rangle(\beta)
   = -\frac{d}{d\beta}\ln\mathcal{Z}$, we evaluate the partition sum
\begin{align}
   \mathcal{Z}
   &= \sum_{i=0}^{M} G(E_{i,IS})\,e^{-\beta E_{i,IS}}.
\end{align}
For clarity, we factorize $\mathcal{Z}=\mathcal{A}\,\mathcal{B}$ into the $i$–dependent contribution
\begin{align}
   \mathcal{A}
   &= \sum_{i=0}^{M}\binom{M}{i}
      \exp\!\left[-\,\frac{2\beta\sigma_K}{\sqrt{M}}\,i\right],
\end{align}
and the remaining prefactor
\begin{align}
   \mathcal{B}
   = \left(\tfrac{1}{2}\right)^M
     \exp\!\left(\beta\sigma_K\sqrt{M}\right).
\end{align}

Using the binomial theorem, $\mathcal{A}$ simplifies to
\begin{align}
   \mathcal{A}
   = \left[1+\exp\!\left(-\tfrac{2\beta\sigma_K}{\sqrt{M}}\right)\right]^M.
\end{align}
Via $\ln\mathcal{Z} = \ln\mathcal{A} + \ln\mathcal{B}$ the derivative can be taken term by term. For $\mathcal{A}$ one obtains
\begin{align}
   -\frac{d}{d\beta}\ln\mathcal{A}
   = \frac{2\sigma_K\sqrt{M}}{\exp\!\left(\tfrac{2\beta\sigma_K}{\sqrt{M}}\right)+1},
\end{align}
and for the prefactor
\begin{align}
   -\frac{d}{d\beta}\ln\mathcal{B}
   = -\sigma_K\sqrt{M}.
\end{align}

Combining both contributions yields the exact expression
\begin{align}
    \langle E_{IS}\rangle(\beta)
    = \sigma_K\sqrt{M}\left[
       -1 + \frac{2}{\exp\!\left(\tfrac{2\beta\sigma_K}{\sqrt{M}}\right)+1}
      \right].
\end{align}

The leading term reproduces the Gaussian result  
\begin{equation}
      \langle E_{IS}\rangle = -\beta\sigma_K^2,
\end{equation}
obtained when $M\to\infty$ or when only the central part of the distribution is sampled. The second term 
${\cal O}(\beta^3/M)$ term represents the deviation from Gaussianity due to the finite low–energy cutoff and grows in relative magnitude upon cooling.

A convenient estimate for the fragile-to-strong crossover temperature $T_{\mathrm{FSC}}$ may be obtained by requiring that the magnitude of the correction reaches a small fraction $c$ of the Gaussian term,
\begin{equation}
   \left|
   \frac{\tfrac{1}{3M}\beta_{\mathrm{FSC}}^{3}\sigma_K^{4}}
        {\beta_{\mathrm{FSC}}\sigma_K^{2}}
   \right| = c,
\end{equation}
leading to
\begin{align}
   T_{\mathrm{FSC}} = \frac{\sigma_K}{\sqrt{M}}\,(3c).
\end{align}
This simple model thus provides an explicit analytical link between the strength of the low-energy depletion (set by $M$) and the temperature scale at which deviations from Gaussian behaviour—and hence the fragile-to-strong crossover—appear.

In the main text, Fig.~\ref{fig:binomial} illustrates these predictions for $M=150$ and $\sigma_K = 1.0$, showing the corresponding behaviour of both $\langle E_{IS}\rangle$ and the apparent activation energy.

\bibliography{FSC}

\end{document}